# The switching of bipolar and unipolar magnetostriction in polycrystalline ZnO film

Suman Guchhait[1], Saumen Chaudhuri[2], and A. K. Das[3,*]

[1]Department of Physics, C. V. Raman Global University, Bhubaneswar, Odisha 752054, India.

[2]Univ Rennes, ENSCR, CNRS, ISCR (Institut des Sciences Chimiques de Rennes), UMR 6226, Rennes, France

[3]Department of Physics, IIT Kharagpur, Kharagpur, West Bengal 721302, India.

*Corresponding author: Dr. Amal Kumar Das.
Electronic mail: amal@phy.iitkgp.ac.in
Telephone: +91-3222-283824. Fax: +91-3222-282700.

**Abstract**

We report the investigation of magnetostrictive property of polycrystalline ZnO film at room temperature on rotating a magnetic field in the plane of the film from 0° to 90° by an indigenously developed optical cantilever beam magnetometer (CBM) setup. In this study, the film exhibits bipolar (tensile & compressive) and unipolar (tensile/compressive) nature of magnetostriction with the angle of rotation of in-plane magnetic field. Moreover, we observe switching behavior of bipolar and unipolar magnetostriction as the applied field is rotated in the plane of the film. The switching of magnetostriction is attributed to the crystal anisotropy of the material. The appearance of bipolar and unipolar magnetostriction has made the ZnO film a suitable candidate for the fabrication of sensors as well as actuators applicable in micro and nano-electronic devices.

## 1. Introduction

The phenomenon of change in dimension, either eleongation or compression, of a magnetic material under the application of magnetic field is referred to as magnetostriction [1-5]. It is the spin-orbit-coupling (SOC) which is responsible for the origin of magnetostriction in the crystals [6]. Depending on the nature of magnetostriction, the application of a magnetostrictive material varies. Usually, a magnetostrictive material with tensile (+) strain is preferable to design actuators, while the material having compressive (-) strain is suitable to fabricate sensors [7-11]. The execution of a magnetostrictive material as an actuator depends on the value of saturation magnetostriction ($\lambda_s$), whereas in case of a sensor, the figure of merit is the strain sensitivity ($d\lambda/dH$) of that particular material [12]. Generally, the rare-earth alloys like $Tb_{0.3}Dy_{0.7}Fe_2$ (Terfenol-D), $SmFe_2$, $TbFe_2$, etc. are known to possess high magnetostriction [13-15]. But the issues like lack of rare-earth elements, high fabrication cost, difficulties in single crystal sysnthesis, brittleness, etc. make this sort of materials to be undesirable to the scientific community [16, 17]. In this context, as an alternate way, we have concentrated on the ZnO to explore its magnetostrictive property mainly because it shows significant room temperature ferromagnetism, excellent mechanical stability, involves low cost and it is easy to sysnthesis [18-21]. ZnO has non-centrosymmetric hexagonal wurtzite crystal structure having a space group of $P6_3mc$ [22, 23]. In this article, we have grown the polycrystalline ZnO film on a cantilever Si substrate by pulsed laser deposition (PLD) system and later investigated the its magnetostrictive property by rotating a bipolar magnetic field of ± 205.62 kA/m in the plane of the film from 0° to 90° with the help of the optical CBM set up [24]. Interestingly, the film is associated with both the bipolar as well as unipolar nature of magnetostriction and also there is switching between the bipolar and unipolar magnetostriction as we rotate the in-plane magnetic field. At small angle, from θ = 15° to θ = 40° the film shows bipolar nature magnetostriction in which up to ~ 100 kA/m the magnetostriction is compressive in nature and beyond 100 kA/m, the magnetostriction is tensile. Now, as we increase the rotation, the bipolar nature gradually decreases and gets switched to unipolar (compressive) nature at θ = 45°. However, on further rotation the unipolar nature of magnetostriction starts to decrease and again it switches back to the bipolar nature at θ = 60° and it persists up to θ = 75°. Beyond this angle, the bipolar nature switches to unipolar (tensile) nature and it carries on up to θ = 90°. So, due to the bipolar nature of magnetostriction in between θ = 15° to θ = 40° and θ = 60° to θ = 75°, the ZnO film can be operated simultaneously as a sensor in the low field region and as an actuator in the high field region. Besides, the film can also be used as a sensor in between θ = 45° to θ = 55° because of the emergence of compressive magnetostriction and as an actuator beyond θ = 75° to θ = 90° due to the arise of tensile magnetostriction.

## 2. Experimental Details

**(a) Structural and Magnetic characteristics:**

The polycrystalline ZnO film was deposited upon the cantilever Si substrate using the PLD technique. The KrF excimer laser source with a wavelength of 248 nm, energy of 400 mJ, and repetition rate of 10

Hz has ablated a well-sintered ZnO target. Nonetheless, a substrate temperature of 500° C, an oxygen pressure of 0.1 mbar, and a base pressure of $3 \times 10^{-5}$ mbar were all maintained throughout the deposition process. We have carried out the X-ray diffraction (XRD) study to investigate the structural information and phase formation associated with the deposited ZnO film. In the XRD study, the film has been scanned from 20° to 70° using a Cu $K_\alpha$ radiation of wavelength 1.54 Å. To identify the smoothness of the film surface, we have captured the two-dimensional (2D) topographical image of the film through the atomic force microscopy (AFM) technique. Besides, we have performed the cross-sectional field emission scanning electron microscope (FESEM) study at a magnification of 50.00 K to measure the thickness of the ZnO film. To investigate the magnetic response of the film, we have carried out the magnetization (M) versus magnetic field (H) measurement at room temperature (300 K) in in-plane and out-of-plane configurations through the superconducting quantum interference device (SQUID).

**(b) Magnetostrictive characteristics:**

We have performed the angle-dependent magnetostriction {λ(θ)} of the polycrystalline ZnO film at room temperature by rotating the in-plane magnetic field from 0° to 90°. The measurement is accomplished with the help of the optical CBM set up under the application of bipolar magnetic field (H) of ± 205.62 kA/m produced through the current source meter, KEPCO 36–28. Here, θ = 0° (along the length of the film) is considered to be the reference angle. By recording the deflection (Δ) of the ZnO/Si composite, the magnetostriction (λ) of the ZnO film at different angles has been calculated with the help of the following equation

$$\lambda(\theta) = \frac{4}{27} \frac{Y_s}{(1+\nu_s)} \frac{(1+\nu_f)}{Y_f} \frac{w_s}{w_f} \frac{t_s^3}{t_f} \frac{\Delta(0) - \Delta(\theta)}{(\beta t_s + t_f)[(l_{eff} - a)^2 - (l_{eff} - b)^2]} \quad (1)$$

The physical significance and the numerical values of each parameter involved in the above equation have been described in detail in our previous section. The rotation geometry of the magnetic field applied in the plane of the ZnO film has been represented schematically in Fig. 1(a). In Fig. 1(b), we have shown the unstrained (H = 0) situation of the ZnO/Si composite, whereas the strained condition (H ≠ 0) of the ZnO/Si composite structure is illustrated in Fig. 1(c).

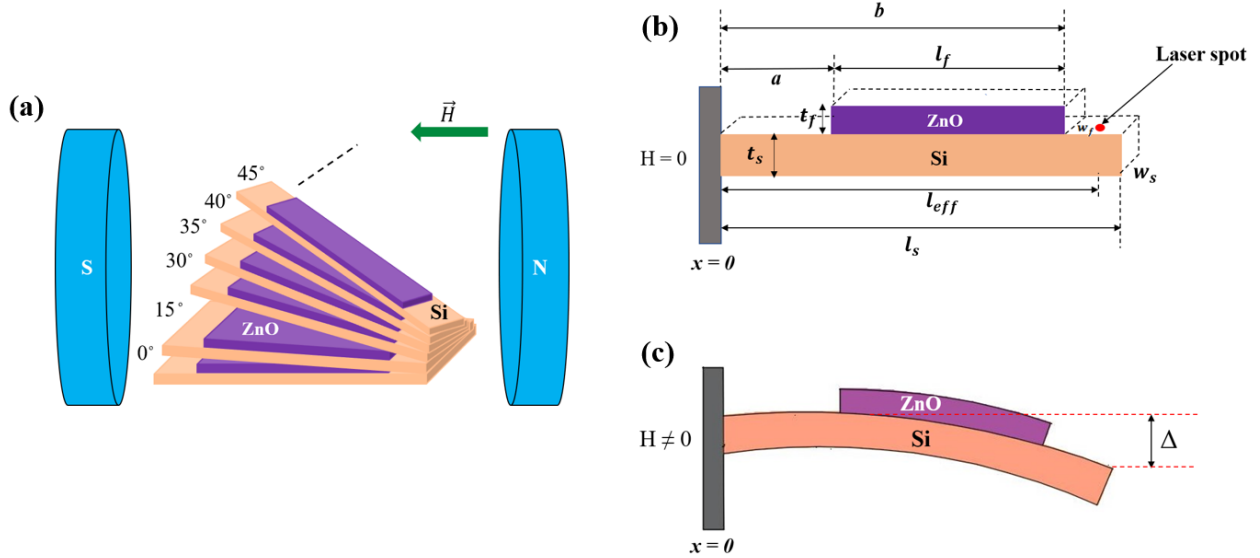

**Fig. 1** (a) Shows the schematic view of the rotation of the magnetic field in the plane of the ZnO film. (b) Shows the unstrained condition of the ZnO/Si composite structure. (c) Represents the schematic view of the ZnO/Si composite in presence of the magnetic field.

## 3. Results and Discussions

**(a) Structural and Magnetic property:**

The hexagonal wurtzite structure of ZnO unit cell is depicted in Fig. 2(a). The XRD spectra of the ZnO/Si composite is depicted in Fig. 2(b). The appearance of extremely high intense (311) peak in the XRD spectra signifies that the growth of the ZnO film is highly c-axis oriented. We have captured the 2D topographical view of the fabricated ZnO film through the AFM study. The film was scanned over (5 μm × 5 μm) surface area and the r.m.s. value of the surface roughness was found to be ~7 nm. The corresponding AFM micrograph is shown in Fig. 2(c). The cross-sectional view of the ZnO/Si composite taken through the FESEM study is represented in Fig. 2(d). From the cross-sectional image, the average thickness of the ZnO film has been found to be ~ 95 nm. The FESEM study also refers to the uniform growth of the ZnO film over the Si substrate. The M-H loops obtained at room temperature in in-plane and out-of-plane configurations are shown in Fig. 2(e). The values of the saturation magnetization are found to be ~ 3000 emu/cc and ~ 1000 emu/cc in in-plane and out-of-plane configurations respectively. The appearance of magnetic order in the deposited ZnO film is attributed to the production of point defects on Zn sites [25, 26]. The defects are typically in the shape of an extended thin layer and are primarily located on the film's surface, at the interface, or close to the interface between the substrate and film [25]. These point defects can be attributed to the deposition of the ZnO film. Now, these point defects can be coupled with spin and hence act as local magnetic moments, resulting in the magnetic response in the ZnO film [27].

**(b) Magnetostrictive property:**

By measuring the deflection of the ZnO/Si composite under the application of magnetic field at different angles, we have calculated the magnetostriction of the ZnO film followed by equation (1). All the magnetostriction vs. magnetic field curves obtained at different angles are shown in Fig. 3 {(a) – (j)}. Here, every (λ – H) loop dictates the butterfly-like hysteretic behavior of magnetostriction with the magnetic field and the presence of arrow marks in each (λ – H) loop indicates the sweep direction of the applied field. The hysteresis nature of every (λ – H) loop is ascribed to the irreversible change of magnetic moments associated with the defects present in the ZnO film. Here, it is evident from the (λ – H) loops that the slope (dλ/dH) is negative up to the field value of (75 – 100) kA/m in all the cases except at θ = 90°. But, beyond 100 kA/m, the slope always carries positive value. Hence, we can argue that the ZnO film exhibits bipolar as well as unipolar magnetostriction and also there is switching between themselves. To understand this, we have taken the maximum values of magnetostriction ($\lambda_{max}$) in the field range, $H_1$ = (75 – 100) kA/m and at the field, $H_2$ = 205.62 kA/m from each (λ – H) loop. The list of the maximum magnetostriction values at $H_1$ and $H_2$ are shown in Table 1. The origin of magnetostrictive response in our fabricated ZnO film is due to the presence of magnetic order.

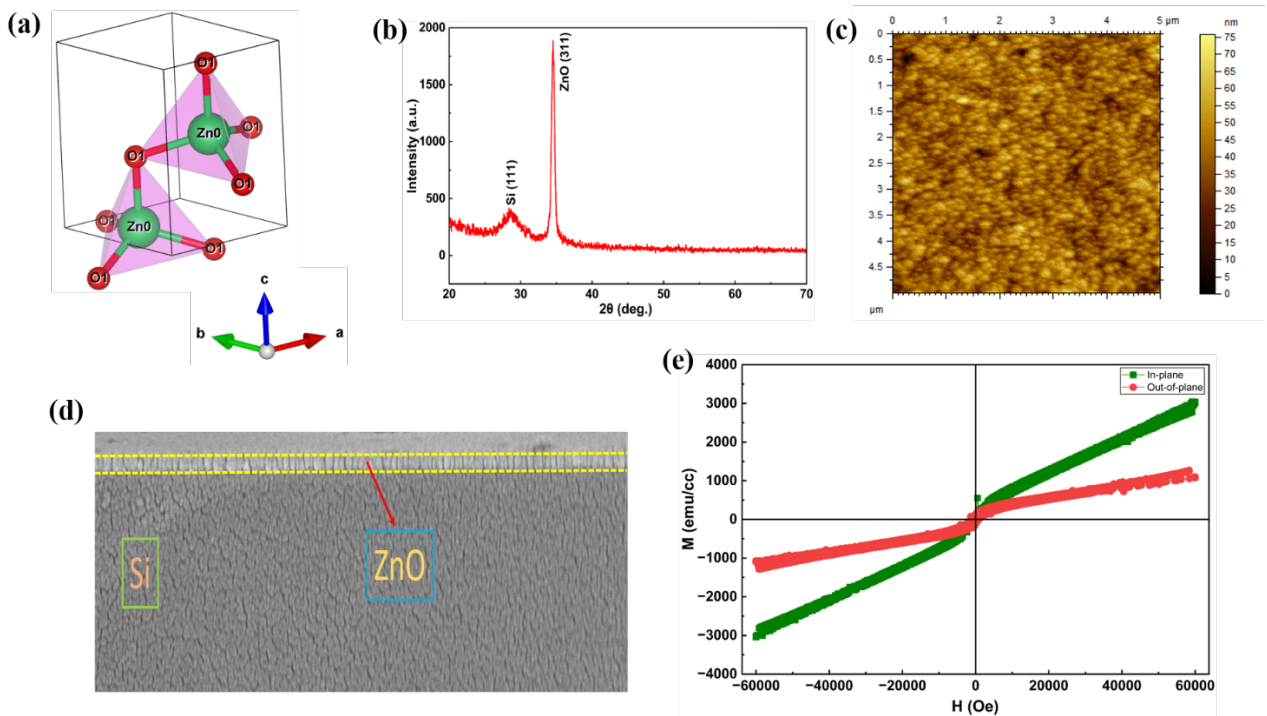

**Fig. 2** (a) Shows the diagram of the unit cell of the ZnO. (b) Represents the XRD spectra of the ZnO/Si composite. (c) Represents the 2D topographical view of the ZnO film obtained through the AFM study. (d) Refers to the cross-sectional FESEM image of the ZnO/Si composite. (e) Highlights the room temperature M-H curves in in-plane as well as out-of-plane geometry obtained through the SQUID magnetometer setup.

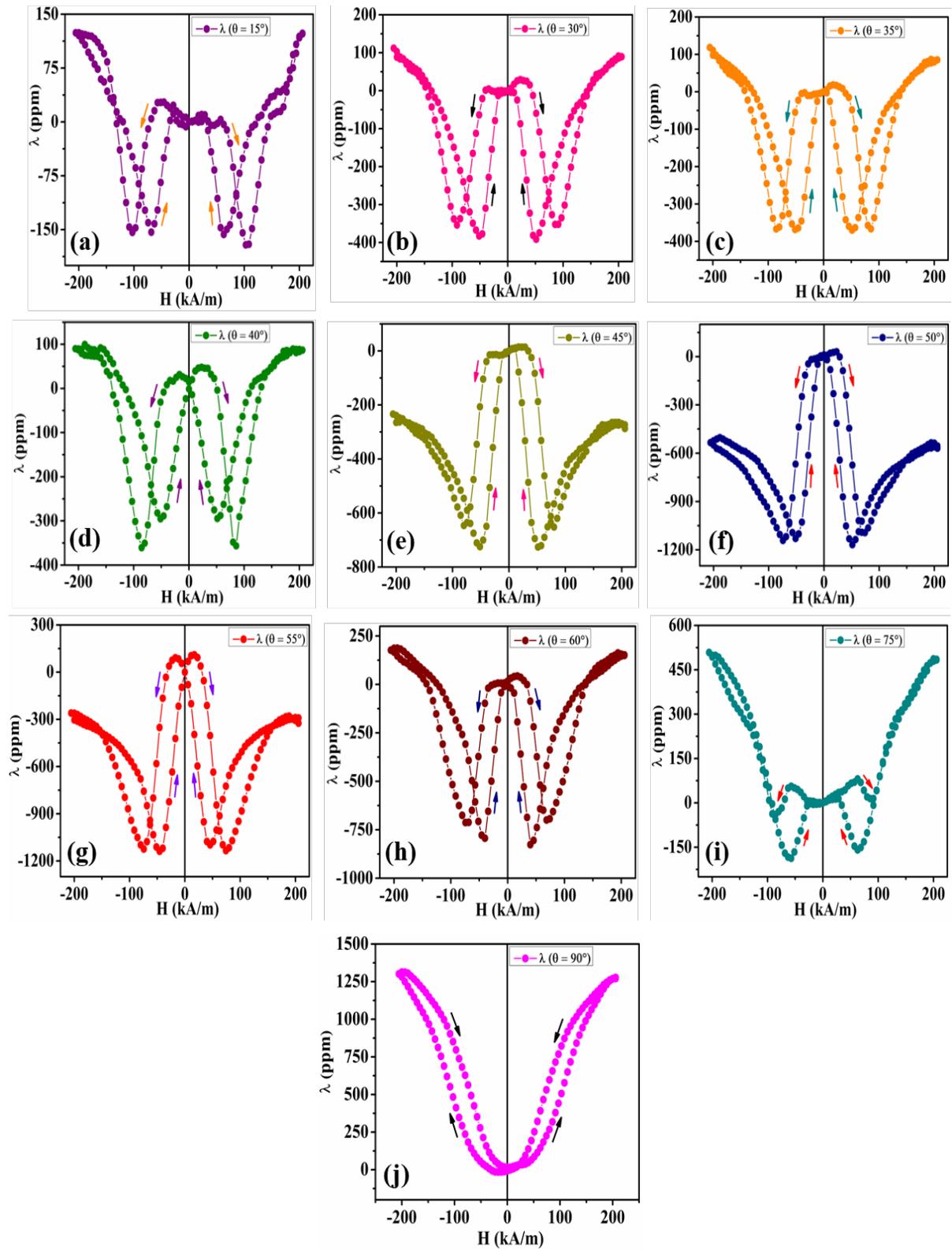

**Fig. 3** {(a) – (j)} Shows the variation of magnetostriction with the magnetic field at various angles.

| Angle (θ) (deg.) | Maximum magnetostriction (ppm) | |
|---|---|---|
| | (75–100) kA/m | 205.62 kA/m |
| 15 | −162.53 | 123.71 |
| 30 | −352.86 | 100.78 |
| 35 | −366.54 | 98.82 |
| 40 | −358.582 | 88.36 |
| 45 | −649.26 | −253.64 |
| 50 | −1117.97 | −562.09 |
| 55 | −1129.09 | −275.88 |
| 60 | −702.92 | 164.45 |
| 75 | −12.43 | 496.77 |
| 90 | 252.00 | 1286.15 |

**Table 1.** Shows the maximum magnetostriction values at different angles in the field range of (75 – 100) kA/m and at 205.62 kA/m.

To understand how does the nature of magnetostriction change with the angle of in-plane magnetic field at $H_1$ and $H_2$, we have made a plot between λ vs. θ and it is represented in Fig. 4(a). According to the Fig. 4(a), the ZnO film exhibits both positive and negative magnetostriction up to the angle, θ = 40°. Therefore, the film is associated with the bipolar (tensile as well as compressive) magnetostriction in the angular range of θ = 15° to θ = 40°. Now, as we increase the angle of rotation above θ = 40°, the positive magnetostriction related to the field, $H_2$ gets switched to negative value at θ = 45° and it remains negative up to θ = 55°. On the other hand, the negative magnetostriction corresponding to the field, $H_1$ also remains negative up to this angle. Hence, in the angular range of θ = 45° to θ = 55°, the ZnO film possesses unipolar (compressive) magnetostriction. On further rotation of the field, the magnetostriction corresponding to the field, $H_2$ decreases and once again switches back to the positive value at $\theta = 60°$ and it remains positive up to $\theta = 90°$. Similarly, beyond $\theta = 55°$, the magnetostriction related to the field, $H_1$ also decays gradually and it finally becomes positive at $\theta = 90°$. So, the film shows the bipolar magnetostriction in the angular range of $\theta = 60°$ to $\theta = 75°$ and unipolar (tensile) magnetostriction at $\theta = 90°$.

From the above discussion, we can summarize that the polycrystalline ZnO film displays bipolar nature of magnetostriction at lower angles. However, as the angle of rotation of the applied magnetic field is increased, the bipolar nature decreases slowly and it becomes unipolar at $\theta = 45°$. On further increment in the rotation, the unipolar nature dies out gradually and it switches back to the bipolar nature once again at $\theta = 60°$. But, as we increase the angle further, the bipolar nature of magnetostriction starts to decay and ultimately gets switched to the unipolar nature at $\theta = 90°$. So, our polycrystalline ZnO film shows bipolar magnetostriction in the angular range of $\theta = 15°$ to $\theta = 40°$ and $\theta = 60°$ to $\theta = 75°$ in which we have observed compressive nature of magnetostriction up to ~ 100 kA/m and tensile nature of magnetostriction above 100 kA/m, whereas it possesses unipolar nature of magnetostriction from $\theta = 45°$ to $\theta = 55°$ and beyond $\theta = 75°$ to $\theta = 90°$. The presence of bipolar and unipolar nature magnetostriction in the ZnO film at different angle of applied field is represented schematically in Fig. 4(b). Depending on the nature of magnetostriction, the polycrystalline ZnO can be used in different ways. As we know, a material having tensile magnetostriction is useful to fabricate different types of actuators, while the material showing compressive magnetostriction is well suited for designing varieties of sensors. So. from Fig. 4(b) we can say that our polycrystalline ZnO film can act simultaneously as a sensor at low magnetic field and as an actuator at high magnetic field when the magnetic field is rotated in the plane of the film from $\theta = 15°$ to $\theta = 40°$ and from $\theta = 60°$ to $\theta = 75°$. Nevertheless, the ZnO film is also applicable to design only sensors in the angular range of $\theta = 45°$ to $\theta = 55°$ due to the origin of negative unipolar magnetostriction, while it can be used to fabricate only actuators beyond $\theta = 75°$ as it shows positive unipolar magnetostriction. Hence, the activity of the ZnO film changes based on the angle of applied magnetic field and this is illustrated schematically in Fig. 4(c).

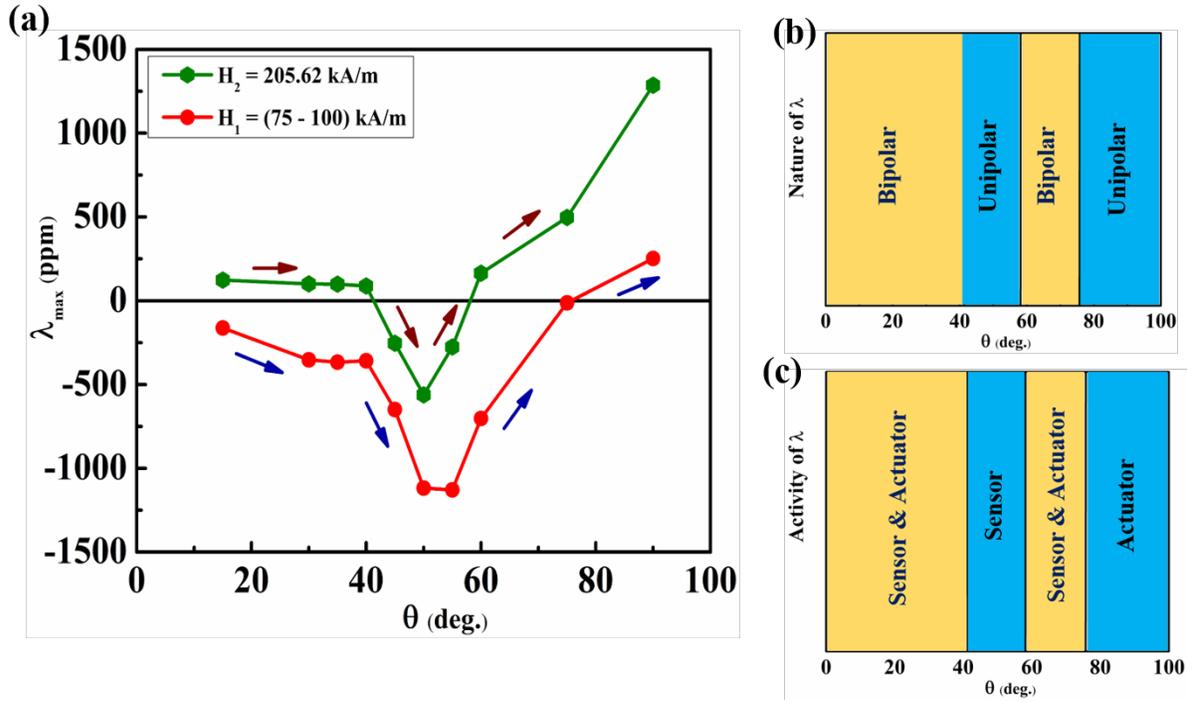

**Fig. 4** (a) Shows the variation of the maximum magnetostriction values obtained in the field range of $H_1$ = 75–100 kA/m and at $H_2$ = 205.62 kA/m with the angle of rotation. (b) Here, we have shown how the nature of the magnetostriction of the ZnO film changes based on the angle of the applied magnetic field. (c) This highlights how the usefulness of the ZnO film varies from angle to angle in the aspect of developing novel magnetostrictive devices.

Let us illustrate the performance of the ZnO film as an actuator and as a sensor more specifically. Already we have mentioned that for the application as an actuator, a material should have high value of positive magnetostriction and to design a sensor, the materials should possess high strain sensitivity (dλ/dH). In Table 2, so far, we have shown that the polycrystalline ZnO film shows remarkably high positive magnetostriction. Therefore, the film is very much suitable for designing different types of actuators based on the magnetostriction phenomenon. Now, to justify the activity of the ZnO film as a sensor, we have additionally calculated the strain sensitivity from the (λ – H) loops appeared from θ = 15° to θ = 75°. To calculate the strain sensitivity, firstly we have chosen the ascending cycle of the (λ – H) loops situated in the 4$^{th}$ quadrant and then taken the (H, λ) values up to the field, 102.81 kA/m. From each (H, λ) value, we have calculated the strain sensitivity and finally plotted them with the magnetic field. The curves showing the variation of dλ/dH with H at the angles from θ = 15° to θ = 75° are represented in Fig. 5{(a) – (i)}. Here, from each dλ/dH vs. H curve, we have identified the maximum value of dλ/dH and listed those values against each angle in table 2. We have found that the values of $(d\lambda/dH)_{max}$ are considerably high at all angles and hence the film is very much suitable for the application as a sensor. However, we have also plotted the $(d\lambda/dH)_{max}$ with the angle. The variation of $(d\lambda/dH)_{max}$ vs. θ has been displayed in Fig. 6. From this curve, we have observed that the values of $(d\lambda/dH)_{max}$ are comparatively lower at small angles. But, as the angle of rotation has been increased, the $(d\lambda/dH)_{max}$ values also increase and attain highest value at an angle, θ = 55°. However, beyond this angle, the values again decrease continuously.

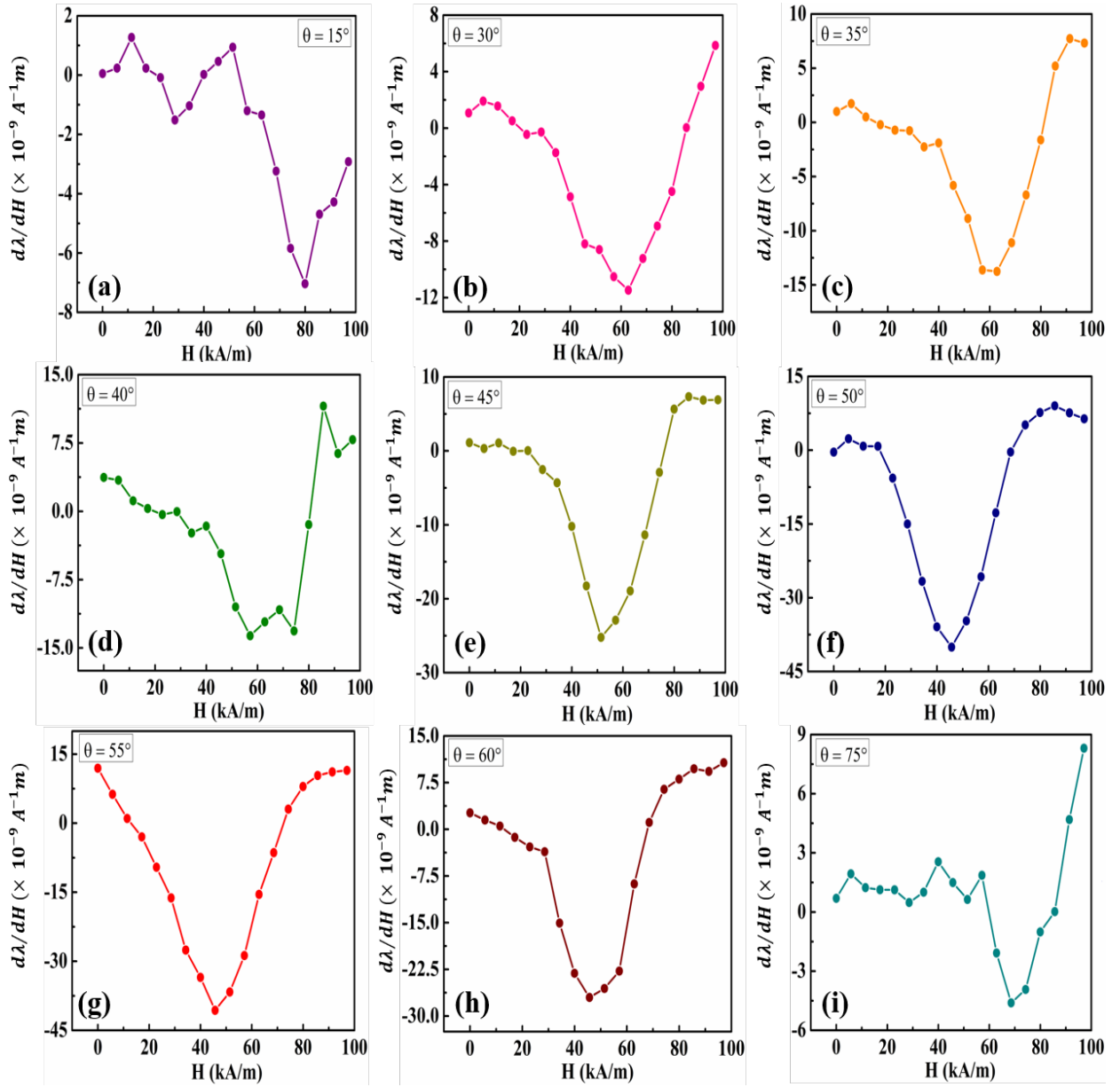

**Fig. 5** {(a) – (i)} Represent the strain sensitivity vs. magnetic field plots in case of ZnO film at various angles.

| θ (deg.) | $(d\lambda/dH)_{max}$ ($\times 10^{-9}$ A$^{-1}$m) |
|---|---|
| 15 | −7.04 |
| 30 | −11.48 |
| 35 | −13.76 |
| 40 | −13.66 |
| 45 | −22.90 |
| 50 | −40.04 |
| 55 | −40.65 |
| 60 | −27.01 |
| 75 | −4.61 |

**Table 2** Shows the numerical values of the maximum strain sensitivities obtained at different angles from the strain sensitivity vs. magnetic field curves.

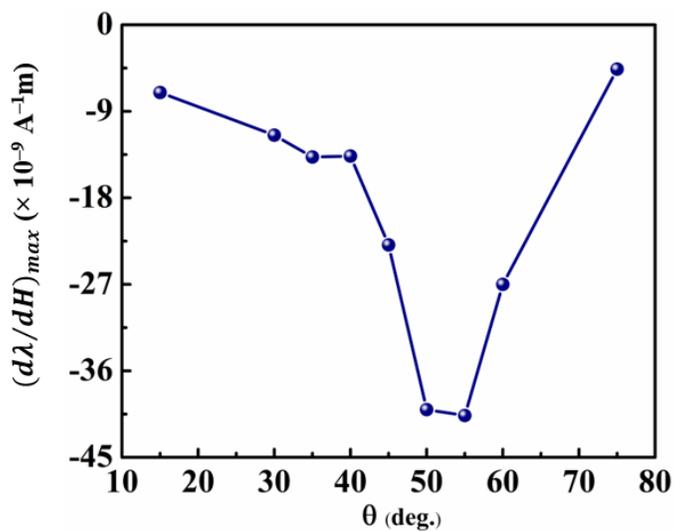

**Fig. 6** Reveals the variation of maximum strain sensitivity with the angle of applied magnetic field.

## 4. Conclusions

In summary, we used PLD system to deposit polycrystalline ZnO film and our indigenously develop optical CBM set up to study the angle-dependent magnetostriction of the fabricated ZnO film at room temperature by rotating the magnetic field in the plane of the film from 0° to 90°. Interestingly, we observed the presence of both bipolar and unipolar nature of magnetostriction in the film and also found the switching between the bipolar and unipolar magnetostriction with the angle of rotation. At lower angles, from $\theta = 15°$ to $\theta = 40°$, the film showed bipolar magnetostriction in which compressive nature dominated at low field strength and tensile nature dominated at high field strength. But, as we rotated the field further, the bipolar nature decreased and it switched to unipolar (compressive) nature at $\theta = 45°$. However, on further increment on the angle of rotation the unipolar nature got diminished and once again switched back to bipolar nature at $\theta = 60°$ and persisted up to $\theta = 75°$. Rotation beyond this angle made the bipolar nature to become unipolar (tensile) once again and it continued up to $\theta = 90°$. Owing to the bipolar nature of magnetostriction in the angular range of $\theta = 15°$ to $\theta = 40°$ and $\theta = 60°$ to $\theta = 75°$, the ZnO film could be used to design sensors in the low field region and actuators in the high field region. Nevertheless, the film could also be applied to develop sensors in between $\theta = 45°$ to $\theta = 55°$ due to the origin of compressive magnetostriction and actuators beyond $\theta = 75°$ to $\theta = 90°$ because of the appearance of tensile magnetostriction.

## References

[1]   Chopra, H. D., & Wuttig, M. (2015). Non-joulian magnetostriction. *Nature*, *521*(7552), 340-343.

[2]   Guchhait, S., Aireddy, H., & Das, A. K. (2021). The emergence of high room temperature in-plane and out-of-plane magnetostriction in polycrystalline CoFe2O4 film. *Scientific Reports*, *11*(1), 22890.

[3]   Fritsch, D., & Ederer, C. (2012). First-principles calculation of magnetoelastic coefficients and magnetostriction in the spinel ferrites CoFe 2 O 4 and NiFe 2 O 4. *Physical Review B*, *86*(1), 014406.

[4]  Kittel, C. (1949). Physical theory of ferromagnetic domains. *Reviews of modern Physics*, *21*(4), 541.

[5]  Guchhait, S., Aireddy, H., Keerthana, Kander, N. S., Biswas, S., & Das, A. K. (2022). Giant magnetostriction and strain sensitivity of ZnFe2O4 film in out-of-plane configuration. *Journal of Applied Physics*, *131*(15), 153903.

[6]  Sukhorukov, Y. P., Telegin, A. V., Bebenin, N. G., Nosov, A. P., Bessonov, V. D., & Buchkevich, A. A. (2017). Strain-magneto-optics of a magnetostrictive ferrimagnetic CoFe2O4. *Solid State Communications*, *263*, 27-30.

[7]  Adhikari, R., Sarkar, A., Limaye, M. V., Kulkarni, S. K., & Das, A. K. (2012). Variation and sign change of magnetostrictive strain as a function of Ni concentration in Ni-substituted $ZnFe_2O_4$ sintered nanoparticles. *Journal of Applied Physics*, *111*(7), 073903.

[8]  Grunwald, A. A. G. O., & Olabi, A. G. (2008). Design of a magnetostrictive (MS) actuator. *Sensors and Actuators A: Physical*, *144*(1), 161-175.

[9]  Hristoforou, E. (2003). Magnetostrictive delay lines: engineering theory and sensing applications. *Measurement Science and Technology*, *14*(2), R15.

[10] Gou, J., Liu, X., Wu, K., Wang, Y., Hu, S., Zhao, H., ... & Yan, M. (2016). Tailoring magnetostriction sign of ferromagnetic composite by increasing magnetic field strength. *Applied Physics Letters*, *109*(8), 082404.

[11] Venkata Siva, K., Sudersan, S., & Arockiarajan, A. (2020). Bipolar magnetostriction in CoFe2O4: Effect of sintering, measurement temperature, and prestress. *Journal of Applied Physics*, *128*(10), 103904.

[12] Zhang, K., Zhang, L., Fu, L., Li, S., Chen, H., & Cheng, Z. Y. (2013). Magnetostrictive resonators as sensors and actuators. *Sensors and Actuators A: Physical*, *200*, 2-10.

[13] Hunter, D., Osborn, W., Wang, K., Kazantseva, N., Hattrick-Simpers, J., Suchoski, R., ... & Takeuchi, I. (2011). Giant magnetostriction in annealed Co1− xFex thin-films. *Nature communications*, *2*(1), 518.

[14]  Samata, H., Fujiwara, N., Nagata, Y., Uchida, T., & Der Lan, M. (1999). Magnetic anisotropy and magnetostriction of SmFe2 crystal. *Journal of Magnetism and Magnetic materials*, *195*(2), 376-383.

[15]  Pateras, A., Harder, R., Manna, S., Kiefer, B., Sandberg, R. L., Trugman, S., ... & Fohtung, E. (2019). Room temperature giant magnetostriction in single-crystal nickel nanowires. *NPG Asia Materials*, *11*(1), 59.

[16]  Wang, H., Zhang, Y. N., Wu, R. Q., Sun, L. Z., Xu, D. S., & Zhang, Z. D. (2013). Understanding strong magnetostriction in $Fe_{100-x}Ga_x$ alloys. *Scientific reports*, *3*(1), 1-5.

[17]  Siva, K. V., Kaviraj, P., & Arockiarajan, A. (2020). Improved room temperature magnetoelectric response in CoFe2O4-BaTiO3 core shell and bipolar magnetostrictive properties in CoFe2O4. *Materials Letters*, *268*, 127623.

[18]  Banerjee, S., Mandal, M., Gayathri, N., & Sardar, M. (2007). Enhancement of ferromagnetism upon thermal annealing in pure ZnO. *Applied Physics Letters*, *91*(18), 182501.

[19]  Hong, N. H., Sakai, J., & Brizé, V. (2007). Observation of ferromagnetism at room temperature in ZnO thin films. *Journal of Physics: Condensed Matter*, *19*(3), 036219.

[20]  Horvath, P., Kiriakidis, G., Nagy, P. M., & Christoulakis, S. (2007, January). Mechanical properties and deformation behavior of ZnO thin films under illumination. In *2007 2nd IEEE International Conference on Nano/Micro Engineered and Molecular Systems* (pp. 421-425). IEEE.

[21]  Alexiadou, M., Kandyla, M., Mousdis, G., & Kompitsas, M. (2017). Pulsed laser deposition of ZnO thin films decorated with Au and Pd nanoparticles with enhanced acetone sensing performance. *Applied Physics A*, *123*, 1-6.

[22]  Carcia, P. F., McLean, R. S., Reilly, M. H., & Nunes Jr, G. (2003). Transparent ZnO thin-film transistor fabricated by rf magnetron sputtering. *Applied Physics Letters*, *82*(7), 1117-1119.

[23]  Konishi, A., Ogawa, T., Fisher, C. A., Kuwabara, A., Shimizu, T., Yasui, S., ... & Moriwake, H. (2016). Mechanism of polarization switching in wurtzite-structured zinc oxide thin films. *Applied Physics Letters*, *109*(10), 102903.

[24] Aireddy, H., & Das, A. K. (2019). The cantilever beam magnetometer for the measurement of electric field controlled magnetic property of ferromagnet/ferroelectrics heterostructures. *Review of Scientific Instruments*, *90*(10), 103905.

[25] Hong, N. H., Sakai, J., & Brizé, V. (2007). Observation of ferromagnetism at room temperature in ZnO thin films. *Journal of Physics: Condensed Matter*, *19*(3), 036219.

[26] Can, M. M., Shah, S. I., Doty, M. F., Haughn, C. R., & Fırat, T. (2012). Electrical and optical properties of point defects in ZnO thin films. *Journal of Physics D: Applied Physics*, *45*(19), 195104.

[27] Zhou, S., & Chen, X. (2019). Defect-induced magnetism in SiC. *Journal of Physics D: Applied Physics*, *52*(39), 393001.
16